\documentclass[%
 apl,%
 amsmath,amssymb,amsfonts,
reprint,%
]{revtex4-1}

\usepackage{graphicx}
\usepackage{epstopdf}
\usepackage{dcolumn}
\usepackage{bm}
\usepackage{natbib}
\usepackage{amstext}
\usepackage{float}
\usepackage[mathlines]{lineno}

\begin{document}

\title{
\begin{center}
Atomistic approach to alloy scattering in $Si_{1-x}Ge_{x}$
\end{center}
} 

\author{Saumitra R Mehrotra}
\email[smehrotr@purdue.edu]

\author{Abhijeet Paul}
\affiliation{Purdue University}

\author{Gerhard Klimeck}
\affiliation{Purdue University}

\begin{abstract}
SiGe alloy scattering is of significant importance with the introduction of strained layers and SiGe channels into CMOS technology. However, alloy scattering has till now been treated in an empirical fashion with a fitting parameter. We present a theoretical model within the atomistic tight-binding representation for treating alloy scattering in SiGe. This approach puts the scattering model on a solid atomistic footing with physical insights. The approach is shown to inherently capture the bulk alloy scattering potential parameters for both n-type and p-type carriers and matches experimental mobility data.
\end{abstract}


\maketitle 


The SiGe alloy has found widespread application in complementary metal-oxide semiconductor (CMOS) technology \cite{Takagi2007}. Continued scaling of CMOS devices has led to the exploration of new materials to augment the silicon channel. Considerable effort has been spent on the realization of ultrathin strained SiGe channel devices \cite{Hoyt2008,Tezuka2003} with alloy scattering being one of the major scattering mechanisms limiting the mobility \cite{Fisch1996,Briggs1998}.

 Alloy scattering has previously been modeled as scattering due to random potential wells of radius R and depth $\Delta U$  \cite{Nordheim1931}. The uncertainity in the definition and large discrepancy in the reported values of $\Delta U$ has widely been discussed in literature \cite{Fisch1996,Briggs1998}. Due to the lack of a consistent physical basis, $\Delta U$ has been relegated to a mere fitting parameter. The work of Fischetti and Laux \cite{Fisch1996} is noteworthy in providing a set of experimentally fitted $\Delta U$ values. There have been some recent theoretical effort in estimating scattering potentials in bulk SiGe using first-principles calculations which render transport calculations computationally burdensome \cite{Joyce2007,Fahy2008}.

Yet further critical questions remain to be tackled.How do alloy scattering rates vary with strain? How does quantum confinement in a system with countable number of atoms affect mobility due to alloy scattering? Can the fitting parameter $\Delta U$ be filled with a physical meaning and insight? To address these questions we consider a fresh look at alloy scattering from an atomistic perspective.

We study the treatment of alloy scattering in SiGe using  a 20 orbital basis set $sp^{3}d^{5}s^{*}$ semi-empirical tight-binding (TB) approach \cite{Klimeck2004}. Previously we developed a virtual crystal approximation (VCA) bandstructure model to study transport in SiGe \cite{Paul2010}. The model was benchmarked against experimental data for bulk band edges and band gaps. Here we analyse an atomistic alloy disorder representation as a perturbation over the homogenous VCA potential. 
\begin{figure}
 \includegraphics{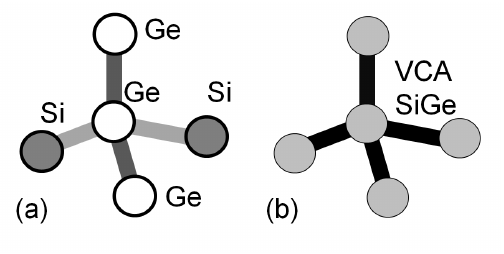}
 \caption{(a) One of the 32 ($N_c=2^5$) possible $Si_{1-x}Ge_{x}$ unit cells in an atomistic representation occuring with probability $p(x)=(1-x)^2x^3$. (b) Unit cell representation in VCA with all bonds and atoms as averaged quantities. }\label{A}
\end{figure}
A VCA TB Hamiltonian for a $Si_{1-x}Ge_{x}$ composition considers ``averaged" atomic species. Each bond length in VCA is assumed to be equal to the averaged Si and Ge bond length for the corresponding composition $x$ (also known as `Vegard limit'). As opposed to VCA an atomistic TB Hamiltonian represents each `atom type' as either Si or Ge and each `bond type' as either Si-Si, Si-Ge or Ge-Ge. In a realistic bulk relaxed SiGe alloy each bond length tries to assume its original bulk bondlength (also  known as `Pauling limit') \cite{Yonenaga2003}. In our calculations, we take each bond to be equal to its bulk bond length while Si-Ge bonds are taken to be half the pure Si and Ge bond lengths. Electronic coupling parameters for the Si-Ge bonds are set to be equal to that of VCA $Si_{0.5}Ge_{0.5}$.

The bulk atomistic Hamiltonian can now be represented as a sum of the VCA Hamiltonian and a disorder term. Within first order perturbation theory the disorder term, $H^{alloy}$ exactly contributes to the alloy scattering.
\begin{eqnarray}
 H^{atomistic}= &&H^{VCA}+ (H^{atomistic}- H^{VCA})\nonumber\\
&& = H^{VCA}+ H^{alloy} 
\label{eq:one}
\end{eqnarray}

 \begin{table*}[t]
\caption{\label{tab:table1}Extracted alloy scattering potentials ${\Delta U_{TB}}(x)$ at the CB and VB edges and compared with experimental values. The last column lists the relative change in $\overline{\Delta U_{TB}}$ with off-diagonal block in Eq. 5 set to zero ($\frac {\lvert\overline{\Delta U_{TB}}-\overline{\Delta U_{TB}}(off diagonal =0)\rvert} {\overline{\Delta U_{TB}}}$), where D,O, and B stand for diagonal, off-diagonal and both. $\overline{\Delta U_{TB}}$ is the best constant fit of ${\Delta U_{TB}(x)}$.}
\begin{ruledtabular}
\begin{center}
\begin{tabular}{ccccc}
 &\multicolumn{3}{c}{Alloy Scattering potential (eV)}&\multicolumn{1}{c}{Main Contributing Block}\\
Band edge ($Si_{1-x}Ge_{x}$) & $\Delta U_{TB}(x)$ & $\overline{\Delta U_{TB}}$ & $\Delta U_{expt/theory*}$ & (relative change)\\ \hline
CB $\Delta$ (x<0.84) &  0.09$x^2$- 0.181$x$+0.963 & 0.90 & 0.7  & B(35$\%$) \\
CB $L$    $(x>0.84)$ & -0.05$x^2$-0.08$x$+1.78      & 1.67 & 1.0,0.6 & O(75$\%$)\\
CB $\Delta-L (x\approxeq0.84)$ &0.914&0.914&-&-\\
VB $\Gamma$ point  & $0.777-0.06x$  &$0.75$ & 0.9,0.81* & D(12$\%$)\\

\end{tabular}
\end{center}
\end{ruledtabular}
\end{table*}

The elastic alloy scattering rate $\tau^{al}_\alpha(E)$ according to Fermi's Golden rule for carriers from a band edge $\alpha$ at wave vector $k_\alpha$ to a band edge $\beta$ at wave vector $k'_\beta$ can now be written as,
\begin{eqnarray}
\frac{1}{\tau^{al}_{\alpha}(E(k_{\alpha}))}=\frac{2\pi}{\hbar} \sum_{k'_\beta} && {\lvert \langle\psi(k_{\alpha}) \lvert H^{alloy}(k_{\alpha}) \rvert \psi(k'_{\beta}) \rangle\rvert}^{2}_{N\times N} \nonumber\\
&&\delta_{E(k_{\alpha}),E(k'_{\beta})} \delta_{\lvert k_{\alpha}\rvert,\lvert k'_{\beta} \rvert}
\end{eqnarray}
 where $N$ is the total number of atoms in the bulk Hamiltonian. The TB wavefunction at band edge $\alpha$  is given by
\begin{eqnarray}
 \psi(k_{\alpha},R)=\frac{1}{\sqrt{N}} \sum_{n\in N} {\sum_{i \in s,p,d,s^{*} } c_{i}\phi_i(R-r_n) \ e^{ik.r_{n}}}
\label{eq:three}
\end{eqnarray}
where $\phi(R-r_{n})$ represents the orthogonalized L\"owdin orbital at site $r_{n}$ ($\alpha,\beta \in \Delta$ or L for the conduction band (CB) and $\Gamma$ for the valence band (VB)).
We assume the case of a perfect random alloy which leads to the condition Eq. 4 \cite{Fahy2008}.
 \begin{eqnarray}
\sum^N_i \sum^N_j \langle H^{alloy}_i \lvert H^{alloy}_j \rangle=\delta_{i,j}
\end{eqnarray}
Using Eq. 3 and 4, Eq. 2 can be written in a condensed form as the sum of scattering due to $N_c$ (=32) different bulk atomistic unit cell types (Figure 1 (a)).
\begin{equation}
\frac{1}{\tau^{al}_{\alpha}(E(k_{\alpha}))}=\frac{2\pi}{\hbar} \Omega_{0}\sum_{\substack{\beta\in N_v \\ j\in N_{c}}} p_{j} (x) {\lvert\langle\psi_{0}(k_{\alpha})\lvert \overline H^j_0 \rvert \psi_{0}(k_{\beta})\rangle\rvert}^2 \rho_\beta(E)
\end{equation}
Here $x$ is the ratio of Ge content, $\Omega_{0}$ ($= a_{0}^3/8$) is the unit cell volume where $a_{0}$ is the $Si_{1-x}Ge_{x}$ lattice constant in Vegard limit. $N_v$ is the number of degenerate valleys which is 6 for $\Delta$, 4 for L and 1 each for heavy and light hole band edge. $\overline H_0^j$ (= $H_0^j$ - $H_0^{VCA}$) is the difference between the $j^{th}$ atomistic unitcell Hamiltonian, $H_0^j$ that occurs with probability $p_j(x)$ and the VCA unitcell Hamiltonian, $H_0^{VCA}$. $\psi_{0}(k_{\alpha})$ is the unit cell wavefunction for band edge $\alpha$ and $\rho_\beta(E)$ is the density of states per spin of the band edge $\beta$ where carriers scatter into ($\alpha$=$\beta$ for intravalley scattering while $\alpha\neq\beta$ for intervalley scattering). Eq. 5 can be written in an expression (Eq. 6) that is commonly used in the literature for alloy scattering rates \cite{Fisch1996}.
 \begin{eqnarray}
\frac{1}{\tau^{alloy}_{\alpha}(E(k_{\alpha}))}=\frac{2\pi}{\hbar} \Omega_{0} && x (1-x) \Delta U^2_{TB}(x) \rho_\beta(E)
\end{eqnarray}

The effective alloy scattering potential  $\Delta U_{TB}(x)$ embedded in the atomistic Hamiltonian can now be calculated using Eq. 5 and 6 as shown in Eq. 7. 
\begin{equation}
\Delta U^2_{TB}(x)= \sum_{\substack{\beta\in N_v \\ j\in N_{c}}} \frac{p_{j}(x)}{x(1-x)}   
{\lvert\langle\psi_{0}(k_{\alpha})\lvert \overline H^j_0\rvert \psi_{0}(k_{\beta})\rangle\rvert}^2
\end{equation}

\begin{figure}[b]
\includegraphics{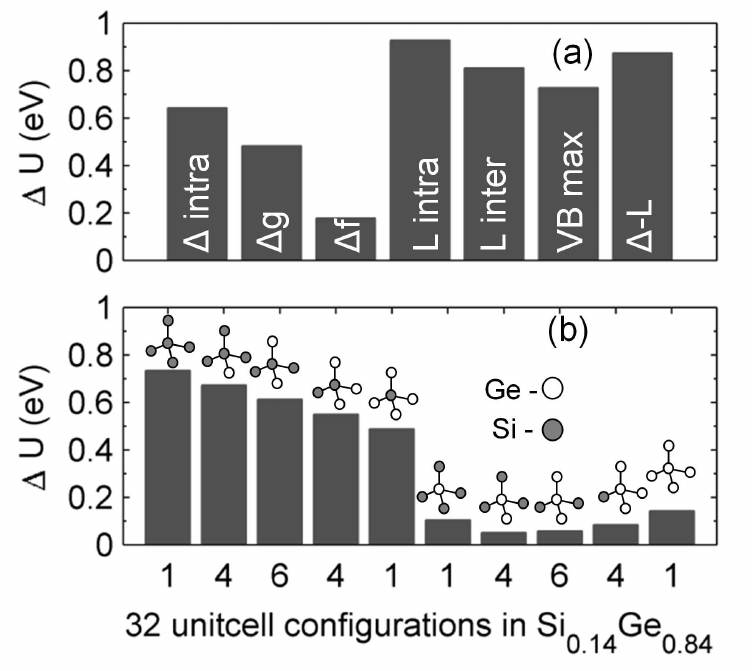}
\caption{(a) Alloy scattering potentials calculated for $Si_{1-x}Ge_{x}$ ($x\backsimeq 0.84$) (b) Calculated transition Matrix elements $\lvert\langle \Psi_{0}(k_\alpha)\lvert H^j_0-H^{VCA}_0 \rvert \Psi_{0}(k_\alpha)\rangle\rvert$ ($\alpha \in \Delta_{[100]}$) in $Si_{0.16}Ge_{0.84}$ due to the 32 possible unit cell configurations contributing to $\Delta$ intravalley scattering.}
\label{fig:B}
\end{figure}

The extracted potentials $\Delta U_{TB}(x)$ are compared with experimentally fitted values in Table ~\ref{tab:table1}. 
The Scattering potential for VB is averaged over all bands at $\Gamma$. A single $\Delta-L$ scattering potential  is quoted which has been averaged between $\Delta$ and L density of states. In the view of the empirical nature of the $\Delta U_{fit}$,the theoretically computed $\Delta U_{TB}$  values are found to be reasonably close. 

Scattering potentials for different intervalley and intravalley transitions are shown in Figure~\ref{fig:B} (a). It is seen that $f-type$ scattering in the $\Delta$ valley is non-zero as opposed to what was assumed in previous calculations \cite{Fisch1996}. Also it is seen that inter-valley scattering is nearly as strong as intra-valley scattering for $L$ minima.  Figure~\ref{fig:B} (b) shows the contributions from different unit cell for $\Delta$ intravalley scattering in bulk $Si_{0.16}Ge_{0.84}$. The scattering potentials are found to depend on both, the atomic species and their neighbourhood. This is in contrast to the standard model which assumes a hard wall potential at each atomic site. Now the understanding of alloy scattering can  be redefined as scattering due to variation in `atom types' and `bond types'. From the last column in Table~\ref{tab:table1} it can be seen that p-type carriers are scattered more due to the variation in `atom type' (Diagonal block) while n-type carriers are scattered more due to variation in `bond type' (Off-diagonal block).

\begin{figure}
 \includegraphics[scale=1]{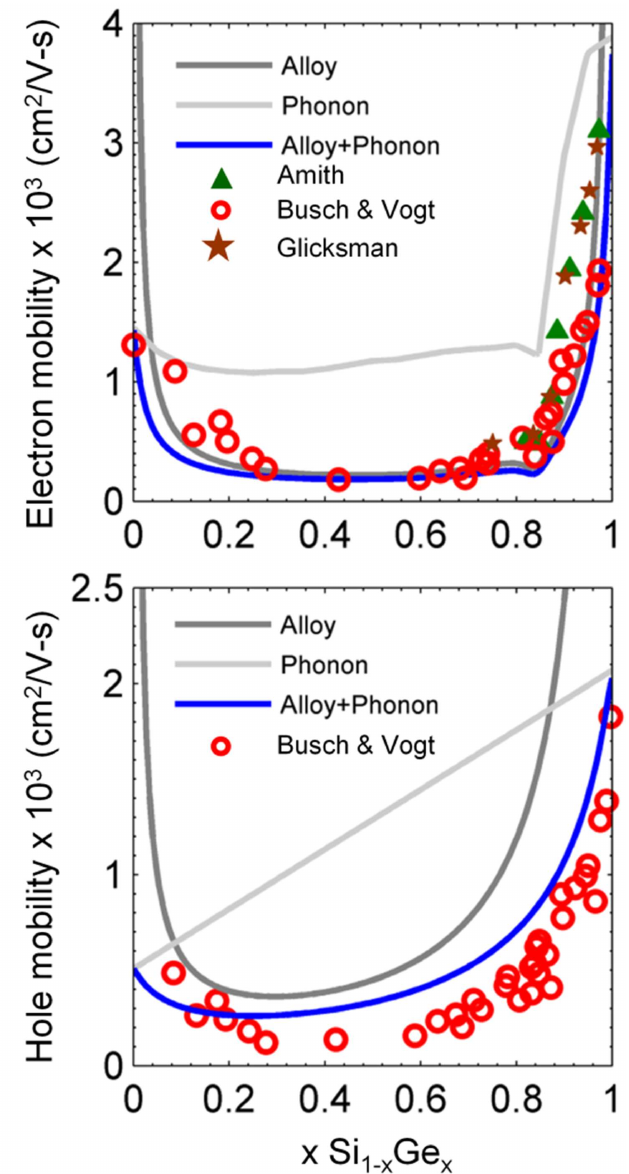}
 \caption{Calculated electron and hole mobility at 300K in bulk relaxed SiGe. Phonon mobility from \cite{Fisch1996}. Experimental data from  $\blacktriangle$ \cite{Amith} $\bigstar$ \cite{glicksman1958mobility} and $\bigcirc$ \cite{busch}.} .
\label{fig:C}
\end{figure}
Our approach to scattering rates is then used to compute bulk mobility for n-type and p-type carriers in $Si_{1-x}Ge_{x}$ and compared to experimental data. We use the Boltzmann transport equation within the isotropic parabolic band approximation similar to \cite{Fahy2008}. Eq. ~\ref{eq:seven} represents the mobility expression for carriers at a conduction band edge $E^\alpha_c$ (limits reversed for VB edge).
\begin{equation}
\mu^{alloy}_{\alpha}=\frac{\frac{4q^2 . N^\alpha_v }{3.m^{VCA}_c kT}  \int_{E^\alpha_c}^{\infty} (E-E^\alpha_c) e^{\frac{E_f-E}{kT}} \rho_{\alpha}(E) \tau^{alloy}_{\alpha}(E) dE }{\frac{qN^\alpha_v}{2\pi^2}(\frac{2m^{VCA}_{dos}}{\hbar^2})^{\frac{3}{2}} \int_{E^\alpha_c}^{\infty}\sqrt{ (E-E^\alpha_c(v))} e^{\frac{E_f-E}{kT}}dE  }
\label{eq:seven}
\end{equation}
 The conductivity ($m^{VCA}_c$) and density of states ($m^{VCA}_{dos}$) masses are extracted from the VCA bandstructure \cite{mehrotra2010}. The final mobility value due to alloy scattering can be calculated using $\mu^{alloy}=\sum_\alpha {n_\alpha \mu_\alpha}/{n}$ where $n_\alpha$ is the carrier density at the band edge $\alpha$ and $n$ is the total carrier density.The bulk SiGe phonon limited mobility $\mu^{phonon}$ is  taken from \cite{Fisch1996}.  Figures 3 (a) and (b) show the computed bulk mobility for electrons and holes respectively. The final mobility is calculated using Matthiessen's rule, $1/\mu =1/\mu^{phonon}+1/\mu^{alloy}. $  A close match is found with the available experimental mobility data for both electrons and holes.

To summarize, we have presented an atomistic approach in a tight-binding representation to treat alloy scattering in SiGe and verified it for bulk against experimental data. The theory predicts $\Delta U$ values which are close to the experimentally fitted numbers. The bulk mobility is computed for both n-type and p-type SiGe materials using the extracted potentials which closely match experimental data. These results indicate the importance of atomistic treatment of alloy scattering in SiGe and further allow studying its effects in nanostructures in a realistic fashion.

Authors would like to thank MSD/MARCO/NSF and Purdue University for financial support. nanoHUB computational resources operated by NCN and funded by NSF are acknowledged.
\nocite{*}
%

 \end{document}